\documentclass[final,5p,times,twocolumn]{elsarticle}

\newcommand\beq{\begin{equation}}
\newcommand\eeq{\end{equation}}
\newcommand{\muR}{\mu_\mathrm{R}}
\newcommand{\muI}{\mu_\mathrm{I}}
\usepackage[normalem]{ulem}  
\usepackage[dvips]{color} 

\renewcommand\sout{\bgroup \color{red} \ULdepth=-.5ex \ULset}

\usepackage{amssymb}
\usepackage{amsmath}

\journal{Physics Letters B}

\begin{document}

\begin{frontmatter}



\title{Topological feature and phase structure of QCD at complex chemical
potential\tnoteref{Report}}
\tnotetext[Report]{Report number: YITP-15-45}

\author[YITP]{Kouji Kashiwa}
\author[YITP]{Akira Ohnishi}

\address[YITP]{Yukawa Institute for Theoretical Physics, Kyoto University,
Kyoto 606-8502, Japan}

\begin{abstract}
The pseudo-critical temperature of the confinement-deconfinement
 transition and the phase transition surface are investigated by using
 the complex chemical potential.
We can interpret the imaginary chemical potential as the Aharonov-Bohm
 phase, then the analogy of the topological order suggests that
the Roberge-Weiss endpoint would define the pseudo-critical temperature.
The behavior of the Roberge-Weiss endpoint at small real quark chemical
 potential is investigated with the perturbative expansion.
The expected QCD phase diagram at complex chemical potential is
 presented.

\end{abstract}


\end{frontmatter}


\section{Introduction}

Understanding the phase structure of Quantum Chromodynamics (QCD) is
one of the most important and interesting subjects
in nuclear and elementary particle physics.
The lattice QCD simulation is a powerful and gauge invariant method,
but it has the sign problem at finite real chemical potential
($\mu_\mathrm{R}$), and we cannot obtain reliable results at large
$\mu_\mathrm{R}$.
Some methods are proposed to circumvent the sign problem, see
Ref.~\cite{deForcrand:2010ys} as an example, but those methods are
limited in the $\mu_\mathrm{R}/T < 1$ region where $T$ is temperature.
Because of the sign problem, low energy effective models of QCD are
extensively used to explore the QCD phase diagram.
Effective models, however, have strong ambiguities and thus
quantitative predictions are impossible at present.
Towards unification of lattice QCD simulations and effective model
approaches, a new method so-called {\it imaginary chemical potential
matching approach}~\cite{Sakai:2008um,Kashiwa:2008bq} is proposed
recently.
In this method, we use lattice QCD data obtained at finite
imaginary chemical potential ($\mu_\mathrm{I}$) to constrain effective
models.
It is well known that the sign problem does not exist in the finite
$\mu_\mathrm{I}$ region and the region has information on the
$\mu_\mathrm{R}$ region; constrained models are reliable not only at
finite $\mu_\mathrm{I}$ but also at finite $\mu_\mathrm{R}$.

In addition to the imaginary chemical potential matching approach,
the concept of the imaginary chemical potential may be important to
define the confinement-deconfinement transition.
In the imaginary time formalism where its time direction is compactified,
the imaginary chemical potential can be interpreted as the
Aharonov-Bohm phase induced by $U(1)$ flux insertions to the
closed time loop.
From this interpretation, we can determine the pseudo-critical
temperature of the deconfinement transition from the Roberge-Weiss (RW)
endpoint~\cite{Roberge:1986mm} with an analogy of the topological
order~\cite{Wen:1989iv,Sato:2007xc} as explained later.

It is also interesting to investigate the phase structure
at {\em finite complex} chemical potential, $\mu=\muR+i\muI$,
where $\muR\not=0$ and $\muI\not=0$.
On the $(T, \muR)$ plane ($\muI=0$),
the first order phase transition may exist,
then we would have the critical point (CP) as the endpoint of the first
order phase transition boundary.
On the $(T, \muI)$ plane ($\muR=0$),
the first order RW transition line exists,
and the RW endpoint can either be the first or the second order.
Some lattice
QCD simulations~\cite{D'Elia:2009qz,Bonati:2010gi,Bonati:2014kpa}
suggest that the order of the RW endpoint is the first order
at the physical pion mass.
Then the RW endpoint is a triple point,
with two other first order lines departing from the RW transition line.
The first order lines appear around the heavy quark limit
as well as around the chiral limit~\cite{D'Elia:2009qz,Bonati:2014kpa}.
These lines with small quark mass may have the chiral transition nature
and are referred to as the {\em chiral critical} lines.
One of the chiral critical lines extending in the $\muI \to 0$ direction
should have an endpoint (chiral critical endpoint; CCE)
before reaching the $\muI=0$
as long as the $\mu=0$ transition is crossover.
\begin{figure}[t]
\begin{center}
 \includegraphics[width=0.45\textwidth]{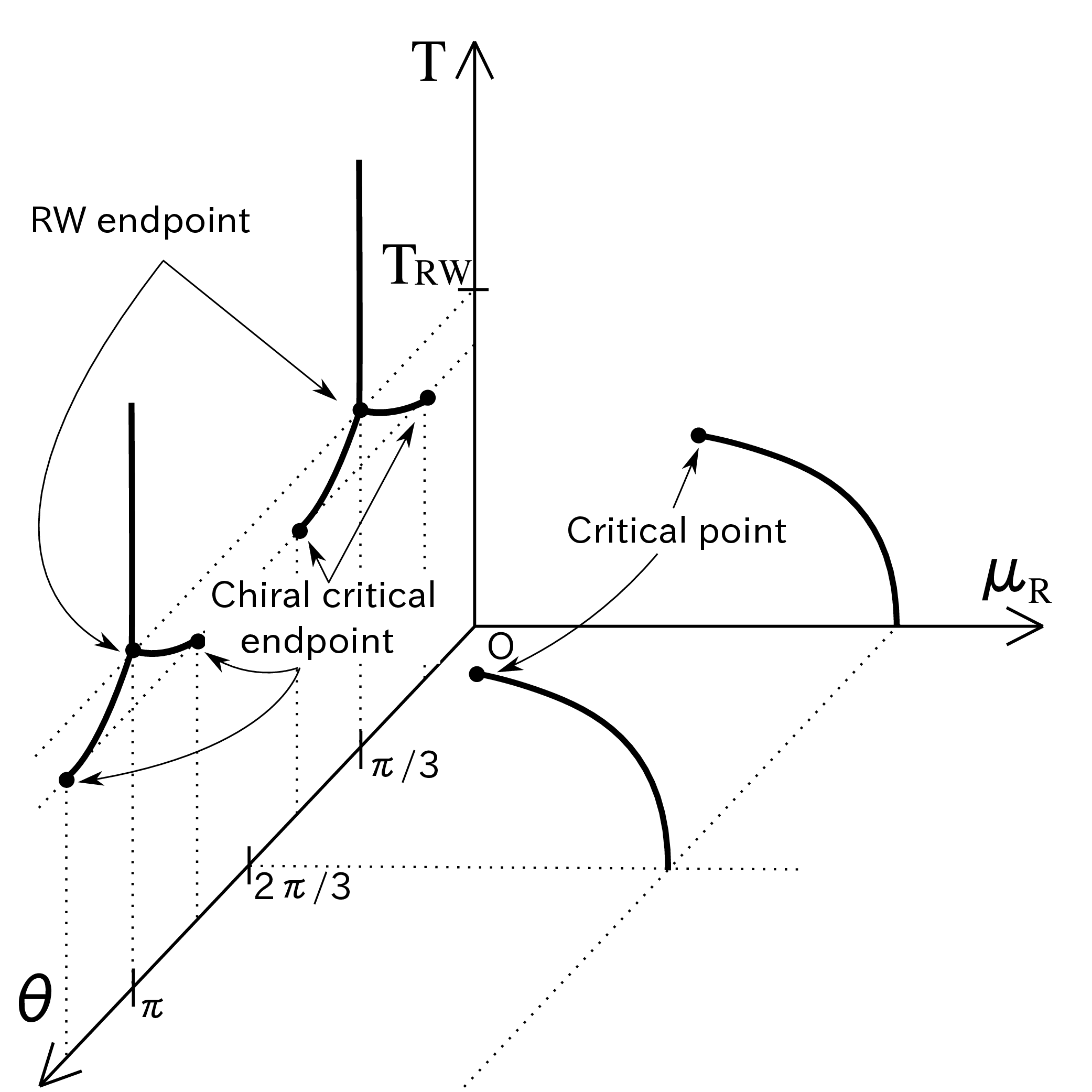}
\end{center}
\caption{
Schematic figure of our current expectation of the QCD phase diagram at
 finite $\mu_\mathrm{R}$ and $\mu_\mathrm{I}$, respectively.
Solid lines represent the first order transition line.
}
\label{Fig:PD-ex}
\end{figure}
Figure~\ref{Fig:PD-ex} summarizes our current expectation of the
QCD phase diagram.
Now we can raise a question.
How does CCE behave at complex chemical potential ?
Specifically, is CCE on the $(T,\muI)$ plane
connected with CP on the $(T,\muR)$ plane
at complex chemical potential, {\em i.e.} in the $(T,\muR,\muI)$ space ?
The topology of the phase diagram at complex chemical potential
would tell us the relation between the deconfinement and the chiral transition.

In this letter, first we briefly summarize properties of QCD
at finite imaginary chemical potential, and propose a new definition of the
pseudo-critical temperature of the deconfinement transition by the
RW endpoint temperature ($T_{RW}$).
Next, we investigate the behavior of the RW endpoint at
small $\mu_\mathrm{R}$ by using the perturbative expansion.
Finally, we present two scenarios of the QCD phase diagram at
complex chemical potential based on the behavior of the RW
endpoint at small $\mu_\mathrm{R}$ and a symmetry argument.

\section{QCD with imaginary chemical potential}

At finite $\mu_\mathrm{I}$, QCD has a special periodicity so-called the
RW periodicity~\cite{Roberge:1986mm}.
The RW periodicity can cause the first order transition lines
(RW transition lines) and their endpoints (RW endpoints).
Those are predicted by using the strong coupling QCD and the perturbative
one-loop effective potential with a background gauge field.
The RW periodicity can be seen from the relation of the grand-canonical
partition function $Z$~\cite{Roberge:1986mm}:
\begin{align}
Z(\theta) &= Z \Bigl( \theta + \frac{2\pi k}{3} \Bigr),
\end{align}
where a dimensionless quark imaginary chemical potential is defined as $\theta
\equiv \mu_\mathrm{I}/T \in \mathbb{R}$ and $k$ is any integer.
If quarks are confined, physical states are classified by hadron degrees
of freedom unlike the deconfined phase.
In the confined and deconfined phases, origins of the RW periodicity are
different:
\begin{description}
\item [Confined phase ---]
      The origin is the dimensionless baryon
      chemical potential $3 \theta$ in the form of $\exp(\pm 3 i
      \theta)$ in the partition function.
      It can be easily seen, for example, from the strong coupling limit
      of the lattice QCD with the mean-field
      approximation~\cite{Nishida:2003fb,Kawamoto:2005mq},
      the chiral perturbation theory with the relativistic Virial
      expansion~\cite{GarciaMartin:2006jj} and that with the finite
      energy sum rule~\cite{Ayala:2011vs}.
\item [Deconfined phase ---]
      The origin is the quark chemical
      potential and the gauge field in the form of
      $\exp[\pm i (g A_4 / T + \theta)]$.
      It generates color non-singlet contributions in the partition
      function.
      It can be understood from the perturbative one-loop effective
      potential with the background gauge
      field~\cite{Gross:1981br,Weiss:1980rj}, and
      the RW periodicity is induced by $\mathbb{Z}_3$
      images~\cite{Roberge:1986mm}.
      Then, we can find the nontrivial degeneracy of the free
      energy minima.
\end{description}
The first order RW transition lines at $\theta = \pi(2k+1)/3$ are
induced by $\mathbb{Z}_3$ images.
Its endpoint which is nothing but the RW endpoint should exist.
The order of the RW endpoint is still under debate, but some lattice
QCD simulations~\cite{D'Elia:2009qz,Bonati:2010gi,Bonati:2014kpa}
suggest that the order seems to be first order around the physical pion
mass.

\section{Imaginary chemical potential and Aharonov-Bohm phase}

In the imaginary time formalism where the temporal direction is compactified,
the $U(1)$ flux can be inserted to the closed loop
of the temporal coordinate.
Then, the imaginary chemical potential can be regarded as the
Aharonov-Bohm phase~\cite{Aharonov:1959fk}.
The Aharonov-Bohm effect has been discussed in the spatial
loop with the flux insertion.
In the imaginary time formalism, the temporal coordinate
shares almost the same features as the spatial coordinates
and thus we can use this interpretation.
In this interpretation, we may use the
discussion of the topological order~\cite{Wen:1989iv}.
Actual applications to zero temperature QCD was discussed in
Ref.~\cite{Sato:2007xc}.
It should be noted that the Polyakov-loop ($\Phi$) is usually
used to discuss the deconfinement transition.
The Polyakov-loop can be obtained from a holonomy
which is the gauge invariant integral along
the temporal coordinate loop.
The Polyakov-loop is the order parameter of the center symmetry
breaking and it is related
with the deconfinement transition in the infinite quark mass limit.
The nontrivial degeneracy of the free energy minima
in the deconfined phase may have some relation
to the complex phase of the Polyakov-loop.
It can be probed by the insertion of the flux as
discussed below.

In Ref.~\cite{Sato:2007xc}, the authors consider the torus $T^3$ at zero
temperature, and introduce three adiabatic operations:
(a) Insert the $U(1)$ flux to spatial closed loops,
(b) exchange $i$-th and $i+1$-th quarks and
(c) move a quark along loops.
Commutation relations of the operation (b) and (c) are described by the
Braid group, and the Aharonov-Bohm effect determines the commutation
relations of those with (a).
If quarks are deconfined, operations become non-commutable because of
the quark's fractional charge.
It is commutable if quarks are confined because physical states are
described by hadron degrees of freedom with integer charges.
Therefore, if there is only one vacuum in the deconfined phase, it is
inconsistent with the non-commutability of the operations and thus
vacuum degeneracy should exist.

At finite temperature, the topological order cannot be well defined,
because thermal states are constructed by a mixture
of pure states with the Boltzmann factor,
and we cannot operate (a), (b) and (c), adiabatically.
Nevertheless,
the RW periodicity shows significantly different behaviors in
the confined and deconfined phases as already mentioned, and it is
induced by the nontrivial appearance of the RW periodicity in the
deconfined phase.
This fact suggests that we can distinguish the confined
and deconfined phases at $\muI=0$ from the
non-trivial degeneracy of the
effective potential at $\theta = \pi/3$.
We here consider the $U(1)$ flux insertion of
$2\pi/3$ to the temporal loop at zero $\muI$.
This corresponds to changing
the $U(1)$ flux from $-\pi/3$ to $\pi/3$
at $\theta=\pi/3$.
These two states are free energy minima at $\theta=\pi/3$
degenerated at
$T>T_\mathrm{RW}$, while they belong to the same minimum at
$T<T_\mathrm{RW}$.
Thus the gluon configurations in states at
$\theta = \pi/3$ are essentially the same as those at $\theta=0$.
This degeneracy seems similar to the vacuum degeneracy in zero
$T$ systems and the analogy can be found;
the response of hidden local minima by the flux insertion via
$\mu_\mathrm{I}$ tells us the non-trivial degeneracy of the free energy
minima.
Therefore, we propose that $T_\mathrm{RW}$ is the pseudo-critical
temperature of the deconfinement transition.

Let us examine if $T_\mathrm{RW}$ provides a deconfinement temperature
reasonably well.
It should be noted that the present definition and the standard
definition determined by using the Polyakov-loop are
consistent in the infinite quark mass
limit where the Polyakov-loop is the exact order-parameter
of the deconfinement transition.
Therefore, we can find the relation
$T_\mathrm{D} = T_\mathrm{\Phi} = T_\mathrm{RW}$
where $T_\mathrm{\Phi}$ is the critical temperature determined by
the susceptibility of the Polyakov-loop and $T_\mathrm{D}$ is
the deconfinement critical temperature.
When the dynamical quark is taken into account,
$T_\mathrm{\Phi}$ becomes the pseudo-critical temperature.
The upper bound of the pseudo-critical temperature can be determined by
using the appearance of local minima of the effective potential as found
in the perturbative one-loop effective potential in
the $\mathrm{Re}~\Phi$ - $\mathrm{Im}~\Phi$ plane.
We call it $T_{\mathbb{Z}_3}$.
From the lattice QCD and effective model predictions, we can have the
relation $T_\mathrm{\Phi} \le T_\mathrm{RW} \le T_{\mathbb{Z}_3}$.
While the Polyakov-loop is no longer the exact order parameter and
thus the determination of $T_\mathrm{D}$ is not unique,
$T_\mathrm{RW}$ is uniquely determined.
Thus, $T_\mathrm{RW}$ which agrees with $T_\mathrm{D}$ in the
infinite quark mass limit is unambiguously determined in the lattice QCD
and effective models and provides a reasonable value as $T_\mathrm{D}$
with the dynamical quark.
If we adopt $T_\mathrm{RW}$ as $T_\mathrm{D}$, we lead to an
implication that the deconfinement transition is the topological phase
transition.

\section{Roberge-Weiss endpoint at complex chemical potential}

We now discuss the $\mu_\mathrm{R}$-dependence of
the pseudo-critical temperature of the deconfinement defined
by $T_\mathrm{RW}$.
We here give an model independent argument based on
the perturbative expansion of the effective potential in
$\mu_\mathrm{R}$ as a first step to investigate the
$\mu_\mathrm{R}$-dependence of the RW endpoint.
It should be noted that non-perturbative model approaches
have several difficulties.
One of the promising effective models is the Polyakov-loop extended
Nambu--Jona-Lasinio (PNJL) model~\cite{Fukushima:2003fw}.
The PNJL model has the model sign problem at finite
$\mu_\mathrm{R}$~\cite{Fukushima:2006uv}.
There are some proposals to circumvent the model sign problem, for
example the complex integral path
contour~\cite{Nishimura:2014rxa,Nishimura:2014kla, Tanizaki:2015pua}
based on the Lefschetz
thimble~\cite{Witten:2010cx,Cristoforetti:2012su,Fujii:2013sra} and
the complex Langevin method~\cite{Parisi:1980ys,Parisi:1984cs}.
Unfortunately those approaches cannot be directly used at finite
complex chemical potential because we cannot maintain the RW
periodicity and some other desirable properties of QCD.

The effective potential at small $\muR$ are expanded
to $\mu_\mathrm{R}^2$ order as
\begin{align}
&{\cal V} (T,\mu_\mathrm{R},\mu_\mathrm{I})
\nonumber\\
&= {\cal V} (T,0,\mu_\mathrm{I})
 - \Bigl( \frac{\mu_\mathrm{R}}{T} \Bigr)
          (T n_q(\mu_\mathrm{R},\mu_\mathrm{I})
          |_{\mu_\mathrm{R}=0} )
\nonumber\\
&
 - \frac{1}{2} \Bigl( \frac{\mu_\mathrm{R}}{T} \Bigr)^2
   \frac{d [ T n_q(\mu_\mathrm{R},\mu_\mathrm{I})]}
        {d \mu_\mathrm{R} / T}
        \Bigl|_{\mu_\mathrm{R}=0}
   +~{\cal O} \Bigl( ( \mu_\mathrm{R}/T)^3 \Bigr)
\label{Eq:perturbation}
\end{align}
where
\begin{align}
T \frac{d n_q}{d \mu_\mathrm{R}/T} \Bigl|_{\mu_\mathrm{R}=0}
&=
T \frac{d n_q}{d (i \mu_\mathrm{I}/T)} \Bigl|_{\mu_\mathrm{R}=0}
= - i T \frac{d n_q}{d \theta} \Bigl|_{\mu_\mathrm{R}=0}.
\label{Eq:21}
\end{align}
Equation~(\ref{Eq:21}) is real, and the second term in r.h.s. of
Eq.~(\ref{Eq:perturbation}) should be pure imaginary.

We here neglect the imaginary part of the effective potential.
This assumption corresponds to the phase quenched approximation.
If the RW endpoint is the weak first order,
the $n_q$ gap at $T_\mathrm{RW}$ is small and the sign problem is mild,
then the phase quenched approximation is justified.
If the RW endpoint is the strong first order,
the partition function at given $(T,\mu)$
is dominated by one classical vacuum, then the role of the imaginary part,
or the phase of the state, is minor.
The phase quenched approximation cannot be applied to the second order
RW endpoint.
Fortunately, the RW endpoint with the realistic quark mass is
predicted to be the first order by lattice QCD
simulations~\cite{D'Elia:2009qz,Bonati:2010gi,Bonati:2014kpa} and
it would be possible to ignore the imaginary part of the effective
potential.
Effects of the imaginary part of the effective potential will be
discussed elsewhere.

Next, we consider the confinement and deconfinement configurations.
We call the configuration at $(T_\mathrm{RW} - \epsilon)$
{\it confinement configuration}
which is labeled as ${\cal C}_{-\epsilon}$ where
$\epsilon$ is a infinitesimal positive value.
Also, we call the configuration at $(T_\mathrm{RW} + \epsilon)$ {\it
deconfinement configuration} which is labeled as
${\cal C}_{+\epsilon}$.

By comparing $\mathrm{Re}~{\cal V}$ with ${\cal
C}_{-\epsilon}$ to that with ${\cal C}_{+\epsilon}$ in the $\epsilon \to
0$ limit,
we can distinguish whether $T_\mathrm{RW}$ decreases or increases
if we can estimate the third term of Eq.~(\ref{Eq:perturbation}).
For example in the lattice QCD and effective model
calculations~\cite{D'Elia:2002gd,Sakai:2008um},
we can see that Eq.~(\ref{Eq:21}) is
negative below $T_\mathrm{RW}$ and it becomes moderate above
$T_\mathrm{RW}$.
The $\mu_\mathrm{R}^2$ correction term makes $\mathrm{Re}~{\cal V}$ with
${\cal C}_{-\epsilon}$ higher than that with $C_{+\epsilon}$ because the
$\mu_\mathrm{R}^2$ correction term is then positive with ${\cal
C}_{+\epsilon}$.
This means that the first-order $T_\mathrm{RW}$ decreases with
increasing $\mu_\mathrm{R}$ at least in the small $\mu_\mathrm{R}$
region.
This behavior is consistent with the decreasing behavior of the
pseudo-critical temperature of deconfinement transition defined by using
usual determinations; for example, see
Refs~\cite{Fukushima:2003fw,Sasaki:2006ww,Kashiwa:2007hw}.

\section{QCD phase diagram at complex chemical potential}

\begin{figure}[tbhp]
\begin{center}
 \includegraphics[width=0.45\textwidth]{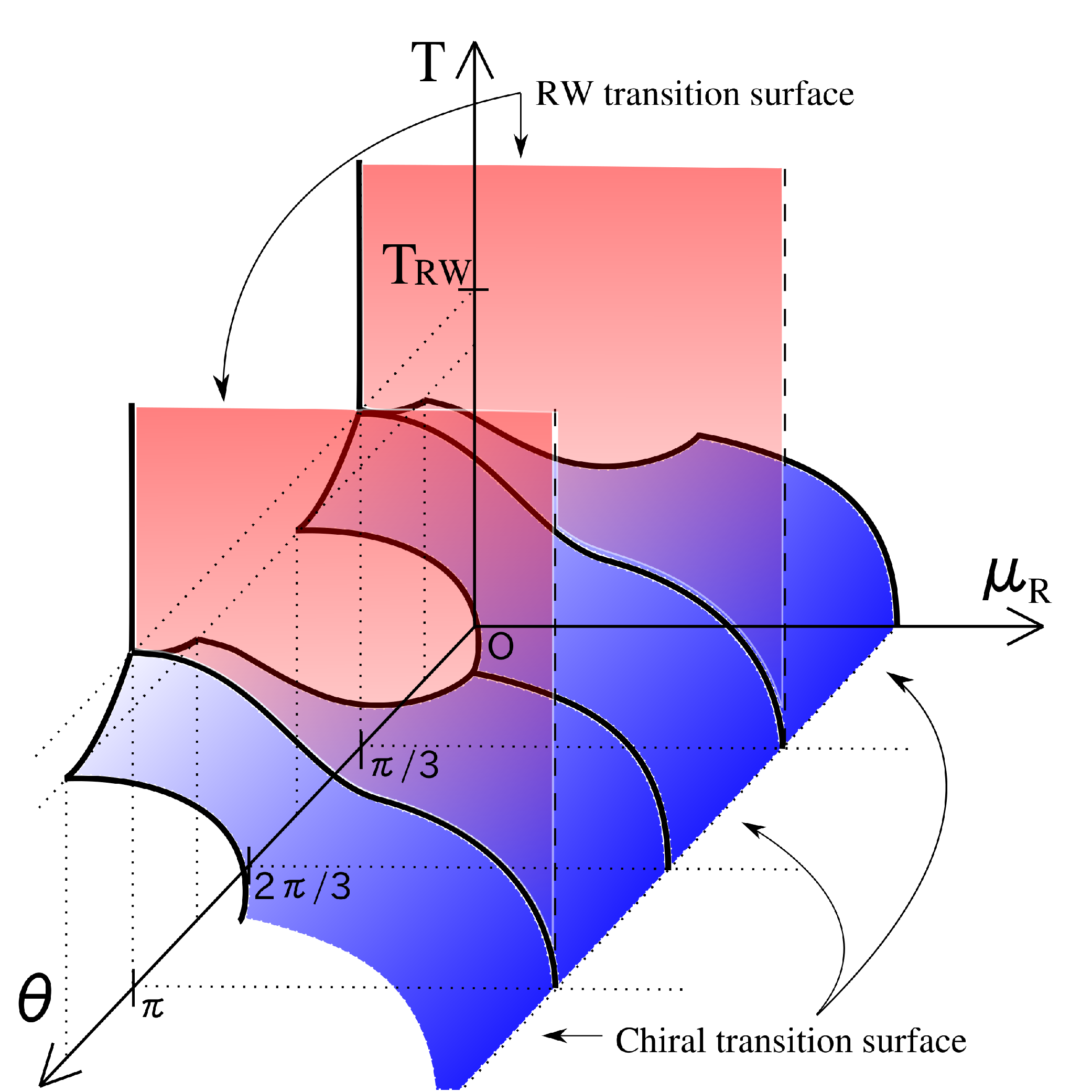}
 \includegraphics[width=0.45\textwidth]{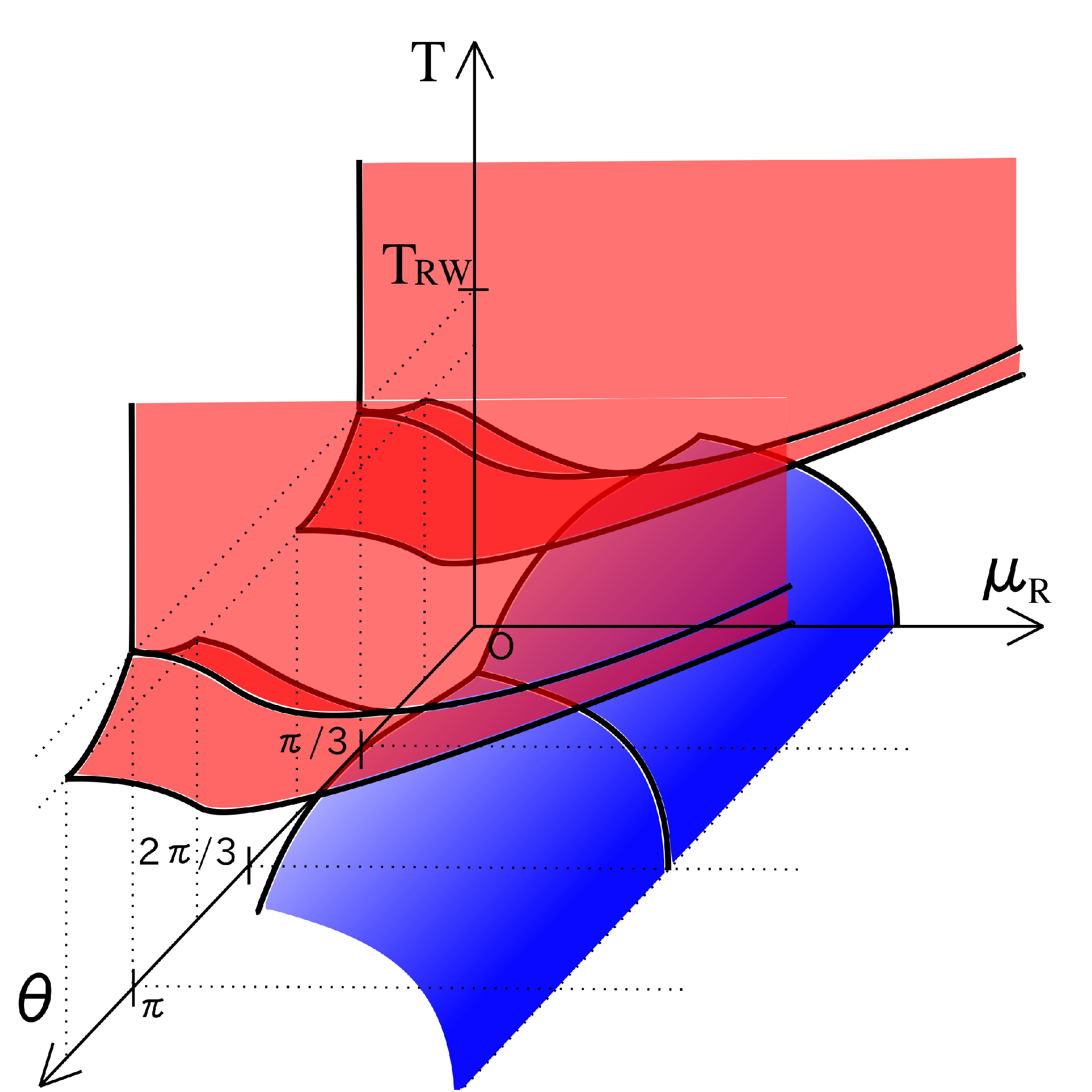}
\end{center}
\caption{
Two possible schematic QCD phase diagrams at finite complex
 chemical potential.
The top (bottom) panel represents the correlated (uncorrelated) case
 between the chiral and RW transition surfaces.
}
\label{Fig:PD-CC}
\end{figure}

By taking into account our perturbative result and the symmetry
argument, we can sketch expected QCD phase diagrams at finite complex
chemical potential.
Phase diagrams expected from our present discussions are
summarized in Fig.~\ref{Fig:PD-CC}.
Because of the RW periodicity, the phase structure should be periodic
along the $\theta$ axis.

The RW transition line on the $(T,\theta)$ plane ($\mu_\mathrm{R}=0$)
may be topologically connected with the first order phase transition
boundary on the $(T,\mu_\mathrm{R})$ plane ($\theta=0$).
Two of the first order transition lines starting from the RW endpoint
have chiral transition nature and are referred to as the {\em chiral
transition lines}~\cite{D'Elia:2009qz}.
Then it is not unreasonable to expect that the endpoint of the chiral
critical line on the $(T,\theta)$ plane is connected with the QCD
critical point on the $(T,\mu_\mathrm{R})$ plane.
In this case, the first order phase boundary on the $(T,\mu_\mathrm{R})$
plane forms a chiral transition surface in the
$(T,\mu_\mathrm{R},\theta)$ space, and connects the $(T,\mu_\mathrm{R})$
plane and the $(T,\theta)$ plane.
The RW transition line extends in the finite $\mu_\mathrm{R}$ region and
forms an RW transition surface in the $(T,\mu_\mathrm{R},\theta)$ space.
The RW endpoint may reach $T=0$ as shown in the top panel of
Fig.~\ref{Fig:PD-CC} or it may deviate
from the chiral transition surface at some temperature.
There is the possibility that $T_\mathrm{RW}$ line becomes smaller than the
chiral critical surface at moderate $\mu_\mathrm{R}$ and finally becomes
larger than the chiral transition surface..
It is deeply related with a strength of the correlation between the
chiral transition surface and the RW transition surface.

Another possibility is that the first order transition lines
on the $(T,\theta)$ plane are topologically separated
from the first order phase boundary on the $(T,\mu_\mathrm{R})$ plane,
as shown in the bottom panel of Fig.~\ref{Fig:PD-CC}.
$T_\mathrm{RW}$ first decreases at small $\mu_\mathrm{R}$, but does not goes
across the chiral transition surface.
In this case, the deconfinement transition represented by the RW
endpoint on the $(T,\theta)$ plane has less
relevance to the
first order phase boundary, which would be the chiral transition, on the
$(T,\mu_\mathrm{R})$ plane.
Therefore, we can call this possibility {\it uncorrelated case} and the
former possibility {\it correlated case}.

The QCD phase diagram at finite complex chemical potential
is related with the following subjects.
(I) In Ref.~\cite{Nakamura:2013ska,Nagata:2014fra}, the authors use
experimental data to construct the canonical partition function.
Then, $T_\mathrm{RW}$ at $\mu_\mathrm{R}=0$ is used to clarify
realized temperatures in experiments through the Lee-Yang zero
analysis~\cite{Yang:1952be,Lee:1952ig}.
In the analysis, $T_\mathrm{RW}$ at finite
$\mu_\mathrm{R}$ should be related with zeros
inside the unit circle on the complex quark fugacity plane. If
so, there is the possibility that we can strictly determine
realized temperatures in experiments if we can systematically
understand the behavior of zeros.
(II) The analytic continuation in QCD from the imaginary to the real
chemical potential is usually performed on the $\mu^2$ plane.
In the continuation, we may miss some information
such as an inhomogeneous condensate~\cite{Kashiwa:2015nya}.
The analytic continuation on the complex chemical potential plane
may restore the information missing.

\section{Summary}

We have proposed that the Roberge-Weiss endpoint provides a
reasonable deconfinement temperature.
The imaginary chemical potential can be interpreted as the Aharonov-Bohm
phase induced by $U(1)$ flux insertions and then the analogy of the
topological order can be used.
In the deconfined phase, we can find the degeneracy of the free energy
by inserting the $U(1)$ flux, and these states belong to different minima.
On the other hand, in the confined phase, we cannot find such
non-trivial structure of the free energy.
This suggests that we can distinguish the confined and deconfined phases
at $\theta=\mu_\mathrm{I}/T=0$ from the non-trivial degeneracy of
the effective potential.
Then, $T_\mathrm{RW}$ becomes the pseudo-critical temperature at
$\mu=0$.

In this study, we have used the analogy of the topological order at
finite $T$.
The topological transition does not have the usual order parameter, but
a relation with an entanglement and topological entropies has been
discussed in the condensed matter physics; for example, see
Ref.~\cite{Levin:2006zz}.
In QCD, such a relation is not clear,
but it is an interesting direction to confirm the topological
nature of the transition.
The calculation of the entanglement entropy in QCD is possible by using
the lattice QCD simulation, and pioneering works were done in
Ref~\cite{Buividovich:2008kq,Nakagawa:2009jk,Nakagawa:2011su}
in the quenched case.
The calculation with the dynamical quark will
provide us information to confirm present discussions.

Using the perturbative expansion in terms of $\mu_\mathrm{R}$, we
have investigated the behavior of $T_\mathrm{RW}$ at small $\mu_\mathrm{R}$
and then the decreasing behavior of $T_\mathrm{RW}$ is obtained.
Based on these results, we presented two scenarios of the QCD phase
diagram at finite complex chemical potential.
First scenario which is the correlated case has the strong correlation
between the chiral and deconfinement transitions.
The RW endpoint at finite $\mu_\mathrm{R}$
finally reaches the $T=0$.
Then, the critical point can become more complex than the usual
expectation since two more first order transition lines can be
connected at the critical point.
The second scenario is the uncorrelated case.
The RW endpoint is
separated from the chiral transition surface.

Since the complex chemical potential is related with the Lee-Yang zero
analysis and the analytic continuation to the finite $\mu_\mathrm{R}$
region, understanding the QCD phase structure at finite complex chemical
potential may have impact to the beam energy scan program in
heavy ion collider experiments, investigation of neutron star structures
and so on.
Our results are based on perturbative calculations and thus the
non-perturbative checks should be done.

K.K. thanks Hiroaki Kouno, Keitaro Nagata and Masanobu Yahiro for useful
 discussions and comments.
He is especially thankful to Yoshimasa Hidaka for his comment which
 motivates us to start this study.
K.K. is supported by Grants-in-Aid for Japan Society for the Promotion
 of Science (JSPS) fellows (No.26-1717).
A.O. is supported in part by KAKENHI
(Nos.
 23340067, 
 24340054, 
 24540271, 
 15K05079, 
 24105001, 
 24105008), 
and by the Yukawa International Program for Quark-Hadron Sciences.




\bibliographystyle{elsarticle-num} 
\bibliography{ref}

\begin{thebibliography}{10}
\expandafter\ifx\csname url\endcsname\relax
  \def\url#1{\texttt{#1}}\fi
\expandafter\ifx\csname urlprefix\endcsname\relax\def\urlprefix{URL }\fi
\expandafter\ifx\csname href\endcsname\relax
  \def\href#1#2{#2} \def\path#1{#1}\fi

\bibitem{deForcrand:2010ys}
P.~de~Forcrand, {Simulating QCD at finite density}, PoS LAT2009 (2009) 010.
\newblock \href {http://arxiv.org/abs/1005.0539} {\path{arXiv:1005.0539}}.

\bibitem{Sakai:2008um}
Y.~Sakai, K.~Kashiwa, H.~Kouno, M.~Yahiro, {Phase diagram in the imaginary
  chemical potential region and extended Z(3) symmetry}, Phys.Rev. D78 (2008)
  036001.
\newblock \href {http://arxiv.org/abs/0803.1902} {\path{arXiv:0803.1902}},
  \href {http://dx.doi.org/10.1103/PhysRevD.78.036001}
  {\path{doi:10.1103/PhysRevD.78.036001}}.

\bibitem{Kashiwa:2008bq}
K.~Kashiwa, M.~Matsuzaki, H.~Kouno, Y.~Sakai, M.~Yahiro, {Meson mass at real
  and imaginary chemical potentials}, Phys.Rev. D79 (2009) 076008.
\newblock \href {http://arxiv.org/abs/0812.4747} {\path{arXiv:0812.4747}},
  \href {http://dx.doi.org/10.1103/PhysRevD.79.076008}
  {\path{doi:10.1103/PhysRevD.79.076008}}.

\bibitem{Roberge:1986mm}
A.~Roberge, N.~Weiss, {Gauge Theories With Imaginary Chemical Potential and the
  Phases of {QCD}}, Nucl.Phys. B275 (1986) 734.
\newblock \href {http://dx.doi.org/10.1016/0550-3213(86)90582-1}
  {\path{doi:10.1016/0550-3213(86)90582-1}}.

\bibitem{Wen:1989iv}
X.~Wen, {Topological Order in Rigid States}, Int.J.Mod.Phys. B4 (1990) 239.
\newblock \href {http://dx.doi.org/10.1142/S0217979290000139}
  {\path{doi:10.1142/S0217979290000139}}.

\bibitem{Sato:2007xc}
M.~Sato, {Topological discrete algebra, ground state degeneracy, and quark
  confinement in QCD}, Phys.Rev. D77 (2008) 045013.
\newblock \href {http://arxiv.org/abs/0705.2476} {\path{arXiv:0705.2476}},
  \href {http://dx.doi.org/10.1103/PhysRevD.77.045013}
  {\path{doi:10.1103/PhysRevD.77.045013}}.

\bibitem{D'Elia:2009qz}
M.~D'Elia, F.~Sanfilippo, {The Order of the Roberge-Weiss endpoint (finite size
  transition) in QCD}, Phys.Rev. D80 (2009) 111501.
\newblock \href {http://arxiv.org/abs/0909.0254} {\path{arXiv:0909.0254}},
  \href {http://dx.doi.org/10.1103/PhysRevD.80.111501}
  {\path{doi:10.1103/PhysRevD.80.111501}}.

\bibitem{Bonati:2010gi}
C.~Bonati, G.~Cossu, M.~D'Elia, F.~Sanfilippo, {The Roberge-Weiss endpoint in
  $N_f$ = 2 QCD}, Phys.Rev. D83 (2011) 054505.
\newblock \href {http://arxiv.org/abs/1011.4515} {\path{arXiv:1011.4515}},
  \href {http://dx.doi.org/10.1103/PhysRevD.83.054505}
  {\path{doi:10.1103/PhysRevD.83.054505}}.

\bibitem{Bonati:2014kpa}
C.~Bonati, P.~de~Forcrand, M.~D'Elia, O.~Philipsen, F.~Sanfilippo, {Chiral
  phase transition in two-flavor QCD from an imaginary chemical potential},
  Phys.Rev. D90~(7) (2014) 074030.
\newblock \href {http://arxiv.org/abs/1408.5086} {\path{arXiv:1408.5086}},
  \href {http://dx.doi.org/10.1103/PhysRevD.90.074030}
  {\path{doi:10.1103/PhysRevD.90.074030}}.

\bibitem{Nishida:2003fb}
Y.~Nishida, {Phase structures of strong coupling lattice QCD with finite baryon
  and isospin density}, Phys.Rev. D69 (2004) 094501.
\newblock \href {http://arxiv.org/abs/hep-ph/0312371}
  {\path{arXiv:hep-ph/0312371}}, \href
  {http://dx.doi.org/10.1103/PhysRevD.69.094501}
  {\path{doi:10.1103/PhysRevD.69.094501}}.

\bibitem{Kawamoto:2005mq}
N.~Kawamoto, K.~Miura, A.~Ohnishi, T.~Ohnuma, {Phase diagram at finite
  temperature and quark density in the strong coupling limit of lattice QCD for
  color SU(3)}, Phys.Rev. D75 (2007) 014502.
\newblock \href {http://arxiv.org/abs/hep-lat/0512023}
  {\path{arXiv:hep-lat/0512023}}, \href
  {http://dx.doi.org/10.1103/PhysRevD.75.014502}
  {\path{doi:10.1103/PhysRevD.75.014502}}.

\bibitem{GarciaMartin:2006jj}
R.~Garcia~Martin, J.~Pelaez, {Chiral condensate thermal evolution at finite
  baryon chemical potential within Chiral Perturbation Theory}, Phys.Rev. D74
  (2006) 096003.
\newblock \href {http://arxiv.org/abs/hep-ph/0608320}
  {\path{arXiv:hep-ph/0608320}}, \href
  {http://dx.doi.org/10.1103/PhysRevD.74.096003}
  {\path{doi:10.1103/PhysRevD.74.096003}}.

\bibitem{Ayala:2011vs}
A.~Ayala, A.~Bashir, C.~Dominguez, E.~Gutierrez, M.~Loewe, et~al., {QCD phase
  diagram from finite energy sum rules}, Phys.Rev. D84 (2011) 056004.
\newblock \href {http://arxiv.org/abs/1106.5155} {\path{arXiv:1106.5155}},
  \href {http://dx.doi.org/10.1103/PhysRevD.84.056004}
  {\path{doi:10.1103/PhysRevD.84.056004}}.

\bibitem{Gross:1981br}
D.~J. Gross, R.~D. Pisarski, L.~G. Yaffe, {QCD and Instantons at Finite
  Temperature}, Rev.Mod.Phys. 53 (1981) 43.
\newblock \href {http://dx.doi.org/10.1103/RevModPhys.53.43}
  {\path{doi:10.1103/RevModPhys.53.43}}.

\bibitem{Weiss:1980rj}
N.~Weiss, {The Effective Potential for the Order Parameter of Gauge Theories at
  Finite Temperature}, Phys.Rev. D24 (1981) 475.
\newblock \href {http://dx.doi.org/10.1103/PhysRevD.24.475}
  {\path{doi:10.1103/PhysRevD.24.475}}.

\bibitem{Aharonov:1959fk}
Y.~Aharonov, D.~Bohm, {Significance of electromagnetic potentials in the
  quantum theory}, Phys.Rev. 115 (1959) 485--491.
\newblock \href {http://dx.doi.org/10.1103/PhysRev.115.485}
  {\path{doi:10.1103/PhysRev.115.485}}.

\bibitem{Fukushima:2003fw}
K.~Fukushima, {Chiral effective model with the Polyakov loop}, Phys.Lett. B591
  (2004) 277--284.
\newblock \href {http://arxiv.org/abs/hep-ph/0310121}
  {\path{arXiv:hep-ph/0310121}}, \href
  {http://dx.doi.org/10.1016/j.physletb.2004.04.027}
  {\path{doi:10.1016/j.physletb.2004.04.027}}.

\bibitem{Fukushima:2006uv}
K.~Fukushima, Y.~Hidaka, {A Model study of the sign problem in the mean-field
  approximation}, Phys.Rev. D75 (2007) 036002.
\newblock \href {http://arxiv.org/abs/hep-ph/0610323}
  {\path{arXiv:hep-ph/0610323}}, \href
  {http://dx.doi.org/10.1103/PhysRevD.75.036002}
  {\path{doi:10.1103/PhysRevD.75.036002}}.

\bibitem{Nishimura:2014rxa}
H.~Nishimura, M.~C. Ogilvie, K.~Pangeni, {Complex saddle points in QCD at
  finite temperature and density}\href {http://arxiv.org/abs/1401.7982}
  {\path{arXiv:1401.7982}}.

\bibitem{Nishimura:2014kla}
H.~Nishimura, M.~C. Ogilvie, K.~Pangeni, {Complex Saddle Points and Disorder
  Lines in QCD at finite temperature and density}\href
  {http://arxiv.org/abs/1411.4959} {\path{arXiv:1411.4959}}.

\bibitem{Tanizaki:2015pua}
Y.~Tanizaki, H.~Nishimura, K.~Kashiwa, {Evading the sign problem in the
  mean-field approximation through Lefschetz-thimble path integral}\href
  {http://arxiv.org/abs/1504.02979} {\path{arXiv:1504.02979}}.

\bibitem{Witten:2010cx}
E.~Witten, {Analytic Continuation Of Chern-Simons Theory} (2010) 347--446\href
  {http://arxiv.org/abs/1001.2933} {\path{arXiv:1001.2933}}.

\bibitem{Cristoforetti:2012su}
M.~Cristoforetti, F.~Di~Renzo, L.~Scorzato, {New approach to the sign problem
  in quantum field theories: High density QCD on a Lefschetz thimble},
  Phys.Rev. D86 (2012) 074506.
\newblock \href {http://arxiv.org/abs/1205.3996} {\path{arXiv:1205.3996}},
  \href {http://dx.doi.org/10.1103/PhysRevD.86.074506}
  {\path{doi:10.1103/PhysRevD.86.074506}}.

\bibitem{Fujii:2013sra}
H.~Fujii, D.~Honda, M.~Kato, Y.~Kikukawa, S.~Komatsu, et~al., {Hybrid Monte
  Carlo on Lefschetz thimbles - A study of the residual sign problem}, JHEP
  1310 (2013) 147.
\newblock \href {http://arxiv.org/abs/1309.4371} {\path{arXiv:1309.4371}},
  \href {http://dx.doi.org/10.1007/JHEP10(2013)147}
  {\path{doi:10.1007/JHEP10(2013)147}}.

\bibitem{Parisi:1980ys}
G.~Parisi, Y.-s. Wu, {Perturbation Theory Without Gauge Fixing}, Sci.Sin. 24
  (1981) 483.

\bibitem{Parisi:1984cs}
G.~Parisi, {ON COMPLEX PROBABILITIES}, Phys.Lett. B131 (1983) 393--395.
\newblock \href {http://dx.doi.org/10.1016/0370-2693(83)90525-7}
  {\path{doi:10.1016/0370-2693(83)90525-7}}.

\bibitem{D'Elia:2002gd}
M.~D'Elia, M.-P. Lombardo, {Finite density QCD via imaginary chemical
  potential}, Phys.Rev. D67 (2003) 014505.
\newblock \href {http://arxiv.org/abs/hep-lat/0209146}
  {\path{arXiv:hep-lat/0209146}}, \href
  {http://dx.doi.org/10.1103/PhysRevD.67.014505}
  {\path{doi:10.1103/PhysRevD.67.014505}}.

\bibitem{Sasaki:2006ww}
C.~Sasaki, B.~Friman, K.~Redlich, {Susceptibilities and the Phase Structure of
  a Chiral Model with Polyakov Loops}, Phys.Rev. D75 (2007) 074013.
\newblock \href {http://arxiv.org/abs/hep-ph/0611147}
  {\path{arXiv:hep-ph/0611147}}, \href
  {http://dx.doi.org/10.1103/PhysRevD.75.074013}
  {\path{doi:10.1103/PhysRevD.75.074013}}.

\bibitem{Kashiwa:2007hw}
K.~Kashiwa, H.~Kouno, M.~Matsuzaki, M.~Yahiro, {Critical endpoint in the
  Polyakov-loop extended NJL model}, Phys.Lett. B662 (2008) 26--32.
\newblock \href {http://arxiv.org/abs/0710.2180} {\path{arXiv:0710.2180}},
  \href {http://dx.doi.org/10.1016/j.physletb.2008.01.075}
  {\path{doi:10.1016/j.physletb.2008.01.075}}.

\bibitem{Nakamura:2013ska}
A.~Nakamura, K.~Nagata, {Probing QCD Phase Structure by Baryon Multiplicity
  Distribution}\href {http://arxiv.org/abs/1305.0760} {\path{arXiv:1305.0760}}.

\bibitem{Nagata:2014fra}
K.~Nagata, K.~Kashiwa, A.~Nakamura, S.~M. Nishigaki, {Lee-Yang zero
  distribution of high temperature QCD and Roberge-Weiss phase transition}\href
  {http://arxiv.org/abs/1410.0783} {\path{arXiv:1410.0783}}.

\bibitem{Yang:1952be}
C.-N. Yang, T.~Lee, {Statistical theory of equations of state and phase
  transitions. 1. Theory of condensation}, Phys.Rev. 87 (1952) 404--409.
\newblock \href {http://dx.doi.org/10.1103/PhysRev.87.404}
  {\path{doi:10.1103/PhysRev.87.404}}.

\bibitem{Lee:1952ig}
T.~Lee, C.-N. Yang, {Statistical theory of equations of state and phase
  transitions. 2. Lattice gas and Ising model}, Phys.Rev. 87 (1952) 410--419.
\newblock \href {http://dx.doi.org/10.1103/PhysRev.87.410}
  {\path{doi:10.1103/PhysRev.87.410}}.

\bibitem{Kashiwa:2015nya}
K.~Kashiwa, T.-G. Lee, K.~Nishiyama, R.~Yoshiike, {Inhomogeneous chiral
  condensates and non-analyticity under an external magnetic field}\href
  {http://arxiv.org/abs/1507.08382} {\path{arXiv:1507.08382}}.

\bibitem{Levin:2006zz}
M.~Levin, X.-G. Wen, {Detecting Topological Order in a Ground State Wave
  Function}, Phys. Rev. Lett. 96 (2006) 110405.
\newblock \href {http://dx.doi.org/10.1103/PhysRevLett.96.110405}
  {\path{doi:10.1103/PhysRevLett.96.110405}}.

\bibitem{Buividovich:2008kq}
P.~V. Buividovich, M.~I. Polikarpov, {Numerical study of entanglement entropy
  in SU(2) lattice gauge theory}, Nucl. Phys. B802 (2008) 458--474.
\newblock \href {http://arxiv.org/abs/0802.4247} {\path{arXiv:0802.4247}},
  \href {http://dx.doi.org/10.1016/j.nuclphysb.2008.04.024}
  {\path{doi:10.1016/j.nuclphysb.2008.04.024}}.

\bibitem{Nakagawa:2009jk}
Y.~Nakagawa, A.~Nakamura, S.~Motoki, V.~I. Zakharov, {Entanglement entropy of
  SU(3) Yang-Mills theory}, PoS LAT2009 (2009) 188.
\newblock \href {http://arxiv.org/abs/0911.2596} {\path{arXiv:0911.2596}}.

\bibitem{Nakagawa:2011su}
Y.~Nakagawa, A.~Nakamura, S.~Motoki, V.~I. Zakharov, {Quantum entanglement in
  SU(3) lattice Yang-Mills theory at zero and finite temperatures}, PoS
  LATTICE2010 (2010) 281.
\newblock \href {http://arxiv.org/abs/1104.1011} {\path{arXiv:1104.1011}}.

\end{thebibliography}

\end{document}